*Stable Outcomes for Two-Sided Contract Choice Problems*
By
*Somdeb Lahiri*
**School of Economic and Business Sciences,**
**University of Witwatersrand at Johannesburg,**
**Private Bag 3, WITS 2050,**
**South Africa.**
**January 2004.**
**(This version: May 7, 2004)**
Email: lahiris@sebs.wits.ac.za
Or lahiri@webmail.co.za

*Abstract*
In this paper we study the cooperative theory of stable outcomes for the room-mates problem modeled as a contract choice problem. We show, that a simple generalization of the Deferred Acceptance Procedure with firms proposing due to Gale and Shapley (1962), yields outcomes for a two-sided contract choice problem, which necessarily belong to the core and are Weakly Pareto Optimal for firms. Under additional assumptions : (a) given any two distinct workers, the set of yields achievable by a firm with the first worker is disjoint from the set of yields achievable by it with the second, and (b) the contract choice problem is pair-wise efficient, we prove that there is no stable outcome at which a firm can get more than what it gets at the unique outcome of our procedure.

1. Introduction :
We consider the problem of choosing a set of multi-party contracts, where each coalition of agents has a non-empty finite set of feasible contracts to choose from. We call such problems, contract choice problems. The economic motivation behind the problem, arises from several real world "commons problems", where agents can pool their initial resources and produce a marketable surplus, which needs to be shared among themselves.

There are clearly, two distinct problems that arise out of such real world possibilities: (i) Coalition Formation: Which are the disjoint coalitions that will form in order to pool in their resources? (ii) Distribution: How will a coalition distribute the surplus within itself? While, the possibility of an aggregate amount of surplus being generated by a coalition is fairly common, there are many situations where more than one aggregate surplus results from a cooperative activity, and the distribution of the surplus depends on the particular aggregate that a coalition chooses to share.

For instance, when a firm employs a worker, it is possible that the revenue generated by the firm depends on whether the worker puts in high or low effort, where high effort corresponds to working full time and low effort corresponds to working half time. The sharing of the surplus between the firm and the worker as well as the amount of the surplus itself, can then depend on the effort put in by the worker and the contingent contract that the firm offers. However, since low effort by the worker, requires more intensive use of the firm's technical capabilities which the firm needs to compensate, the firm would in such a situation retain a greater proportion of the surplus. As an example, consider a situation where a high effort by the worker allows him 60% of the surplus, whereas a low effort allows him only 30%. If high effort by the worker yields a surplus of dollar 100, and low effort yields a surplus of dollar 90, then with high effort, the firm retains dollar 40 and the worker gets dollar 60, whereas with low effort the firm retains dollar 63 and the worker gets dollar 27.

In our model each non-empty subset of agents has a non-empty finite set of pay-off vectors to choose from. An outcome comprises a partition of the set of agents, and an assignment for each coalition in the partition a feasible pay-off vector. Our model is therefore a special kind of cooperative game with non-transferable utility. A stable outcome is an outcome that is not blocked by any coalition, i.e. an outcome that no coalition can improve upon. Thus, the set of stable outcomes corresponds to the core of the corresponding cooperative game.

A salient feature of many markets is to match one agent with another. This is particularly true, in the case of assigning workers to firms. Such markets are usually studied with the help of "two sided matching markets" introduced by Gale and Shapley (1962). However, not all matching problems where disjoint pairs are required to form, are dichotomous. For instance in a doubles version of a tennis tournament, pairs are formed from a given pool of players, without any obvious dichotomy existing within the pool. The problem of forming disjoint pairs out of a given set of agents is what Gale and Shapley(1962) called a room-mates problem. A room-mates problem corresponds to the formation of partnerships between pairs of agents as is observed for instance in the case of firms providing legal or accounting services. The two-sided matching market of Gale and Shapley (1962) (with one side of the market comprising workers and the other side comprising firms), is indeed a special case of the room-mates problem. The solution concept proposed by Gale and Shapley (1962), called a stable matching, requires that there should not exist two agents, who prefer each other, to the individual they have been paired with. It was shown in Gale and Shapley (1962), in a framework where every agent has preference defined by a linear order over the entire set of agents, that a room-mates problem may not admit any stable matching although a two-sided matching market always does. Indeed, given a two-sided matching market, there is always a stable matching which no firm considers inferior to any other stable matching, and there is always a stable matching that no worker considers inferior to any other stable matching. An overview of the considerable literature on two-sided matching markets that has evolved out of the work of Gale and Shapley (1962), is available in Roth and Sotomayor (1990).

In this paper, we represent a generalization of the room-mates problem as a contract choice problem, where the coalition structure in any outcome comprises either one or two element sets. The first significant result that we present here is a strengthening of a result originally due to Knuth (1976). The original result said that given any two stable matchings for a two-sided contract choice problem, if no firm prefers the first matching to the second, then no worker prefers the second matching to the first. In our framework of a room-mates problem represented as a contract choice problem we show the following:

**Given two outcomes for a room-mates problem of which say the second is stable, and given a non-empty subset of agents S if (a) every agent in S prefers the second outcome to the first, and (b) no agent in S and his room-mate under the second matching prefer each other to their respective room-mates in the first matching, then no room-mate of an agent in S prefers the second matching to the first.**

A particular instance of a generalized room-mates problem is a two-sided contract choice problem, which however is a generalization of the model due to Eriksson and Karlander (1998). We allow each pair of agents a non-empty finite set of real valued divisions of a good to choose from. Each agent is assumed to prefer more of the good to less of it. Further, the set of agents are divided into two disjoint sets, with one set being the set of firms and the other the set of workers, with no pair of agents on the same side of the market being able to obtain an allocation which is at least as good as an allocation that could be obtained by them remaining single or by forming a pair with a member on the other side of the market. If each pair of agents is provided singletons to choose from, then we have the two-sided matching market of Gale and Shapley (1962).

We show here, that a two-sided contract choice problem invariably admits a non-empty core. A simple generalization of the Deferred Acceptance Procedure with firms making offers due to Gale and Shapley (1962), yields outcomes for the two-sided contract choice problem, which necessarily belong to the core. The main difference between the procedure we define and the Deferred Acceptance Procedure, is that a firm can make offers to the same worker several times.

We also show, that any outcome of this procedure is Weakly Pareto Optimal for Firms, i.e. there is no other outcome which all firms prefer to an outcome of this procedure. This result is an extension to our framework, of a similar result due to Roth and Sotomayor (1990). Under the additional assumptions: (a) given any two distinct workers, the set of yields achievable by a firm with the first worker is disjoint from the set of yields achievable by it with the second, and (b) the contract choice problem is pair-wise efficient, we prove that the procedure has a unique outcome. Further, there is no stable outcome at which a firm can get more than what it gets at this unique outcome.

As in Sotomayor (1996), it is possible to provide a non-constructive proof of the existence of a stable outcome, in the framework of a two-sided contract choice problem. Such a proof is essentially non-algorithmic although as Sotomayor (1996) shows, is much simpler than its procedural counterpart. A consequence of such a proof is the absence of an explicit "design" for a stable outcome.

2. The Model : Let X be a non-empty finite subset of $\aleph$ (: the set of natural numbers), denoting the set of agents. We assume that each agent prefers more money to less.

Let $\Re$ denote the set of all real numbers and $\Re_+$ the set of non-negative real numbers. Let [X] denote the set of all non-empty subsets of X. Members of [X] are called coalitions. Given $S \in [X]$, let $\Re^S$ denote the set of all functions from S to $\Re$ and $[\Re^S]$ denote the set of all non-empty finite subsets of $\Re^S$.

A Contract Choice Problem (CCP) is a function G: $[X] \to \bigcup_{S \in [X]} [\Re^S] \cup \{\phi\}$, such that for all (i) for all $S \in [X]$: $G(S) \subset \Re^S$; (b) for all $a \in X$: $G(\{a\}) = \{0\}$.

For $S \in [X]$, G(S) is the set of all feasible allocations of money for agents in S. If G(S) is empty then the set of feasible allocations for S is empty. Hence such a coalition S cannot materialize.

A CCP G is said to be super-additive if for all $S, T \in [X]$, with $S \cap T = \phi$ : [$x \in G(S)$, $y \in G(T)$ and $G(S \cup T) \neq \phi$ ] implies [$z \in G(S \cup T)$, where z(a) = x(a) for all $a \in S$ and z(a) = y(a) for all $y \in T$].

Given a CCP G, a coalition structure for G is a partition of X.

A pay-off function is a function v : $X \to \Re_+$. If v is a pay-off function and $S \in [X]$, then v|S denotes the restriction of v to the set S.

An outcome for a CCP G is a pair (h, v), where h is a coalition structure for G and v is a pay-off function such that (i) for all $a \in X$: $v(a) \geq 0$; (ii) for all $S \in h$: $v|S \in G(S)$.

The pair (h, v), where h = {{a}/ $a \in X$} and v(a) = 0 for all $a \in X$, is an outcome for every CCP. Hence the set of outcomes is always non-empty.

A special case of a CCP is a generalization of the room-mates problem of Gale and Shapley (1962), where $G(S) = \phi$, whenever #S > 2. We call such problems generalized room-mates problems.

Given a coalition structure h for a generalized room-mates problem G, we can define one to one function $\mu: X \to X$, such that for all $a \in X$: {a, $\mu(a)$} $\in h$. $\mu$ is called the matching associated with h and satisfies the property [for all $a \in X$: $\mu(\mu(a)) = a$].

Let Y = {$S \in [X]$/ #S = 2} i.e. coalitions of size two.

*Given an outcome (h, v) for a generalized room-mates problem G, a coalition $S \in Y$ is said to block (h, v) if there exists $x \in G(S)$: x(a) > v(a) for all $a \in S$.*
*An outcome (h, v) for a generalized room-mates problem G is said to belong to the core of G, if it does not admit any blocking coalition. Let Core(G) denote the set of outcomes in the core of G. An outcome in Core(G) is said to be a stable(stable outcome) for(of) G.*

The following example due to Gale and Shapley (1962) shows that the core of a generalized room-mates problem may be empty.

Example 1 (Gale Shapley (1962)) : Let X = {1,2,3,4}. For $a \in X$, let $u^a: X \to \Re$ be defined as follows:
$u^1$: $u^1(2) = 3$, $u^1(3) = 2$, $u^1(4) = 1$, $u^1(1) = 0$;
$u^2$: $u^2(3) = 3$, $u^2(1) = 2$, $u^2(4) = 1$, $u^2(2) = 0$;
$u^3$: $u^3(1) = 3$, $u^3(2) = 2$, $u^3(4) = 1$, $u^3(3) = 0$;

$u^4$: $u^4(1) = 3$, $u^4(2) = 2$, $u^4(3) = 1$, $u^4(4) = 0$.

$u^a(b)$ is the worth to agent a of forming a coalition/ partnership with agent b. Let, G be the generalized room-mates problem such that for all $S \in [X]$: (i)$G(S) = \{x \in \Re^S /$ either $[x(a) = u^a(b)$ and $x(b) = u^b(a)]$ or $[x(a) = u^a(a)$ and $x(b) = u^b(b)]\}$ if $S = \{a,b\}$, $a,b \in X$; (ii) $G(S) = \phi$ if $\#S > 2$.

Thus, for coalitions $S = \{a,b\}$ of size 2, the feasible yields comprise the one that result from the two agents in the coalition actively collaborating $[x(a) = u^a(b)$ and $x(b) = u^b(a)]$ as well as the one that results when they decide to remain single $[x(a) = u^a(a)$ and $x(b) = u^b(b)]$. The latter yield vector has been included in order to obtain a super-additive generalized room-mates problem, and could be easily omitted without affecting any of our conclusions.

Suppose (h,v) is an outcome such that $v(4) \neq 0$. If $v(4) = 1$, then $\{3,4\} \in h$ and $v(3) = 1$. Thus, $\{2,3\}$ blocks (h,v), since 2 can get 3 units and 3 can get 2 units in $G(\{2,3\})$; if $v(4) = 2$, then $\{2,4\} \in h$ and $v(2) = 1$. Thus, $\{1,2\}$ blocks (h,v) since 1 can get 3 units and 2 can get 2 units in $G(\{1,2\})$; if $v(4) = 3$, then $\{1,4\} \in h$ and $v(1) = 1$. Thus, $\{1,3\}$ blocks (h,v) since 3 can get 3 units and 1 can get 2 units in $G(\{1,3\})$. Thus, $v(4) \neq 0$ implies (h,v) does not belong to Core(G). Hence suppose $v(4) = 0$. If $v(3) = 0$, then both $\{2,3\}$ and $\{3,4\}$ block (h,v); if $v(2) = 0$, then both $\{1,2\}$ and $\{2,4\}$ block (h,v); if $v(1) = 0$, then both $\{1,3\}$ and $\{1,4\}$ block (h,v). Since $v(4) = 0$ requires $v(a) = a$ for at least one $a \in \{1,2,3\}$, Core(G) = $\phi$.
It is worth observing that G is a super-additive CCP, since $G(S) = \phi$ if $\#S > 2$.

3. Properties of Stable Outcomes for Room-mates Problems:

Lemma 1: Let (h,v) and (h',v') be stable outcomes for a generalized room-mates problem G. Let $\{a,b\} \in h$ and $b \neq a$. If $v(a) > v'(a)$, then $v'(b) > v(b)$.

Proof: Suppose (h,v), (h',v'), a and b are as above. If $v(a) > v'(a)$, then the stability of (h',v') requires $v'(b) \geq v(b)$. Q.E.D.

The following lemma extends one due to Knuth (1976):

Lemma 2: Let G be a generalized room-mates problem for which (h',v') is a stable outcome. Let S be a non-empty subset of X. Suppose that (h,v) is an outcome such that $v'(a) > v(a)$ for all $a \in S$. If there does not exist $a \in S$, such that $\{a,b\} \in h'$ blocks (h,v), then $v(b) \geq v'(b)$ whenever $b \in \mu'(S)$, where $\mu'$ is the matching associated with h'.

Proof: Suppose that (h,v) is an outcome such that $v'(a) > v(a)$ for all $a \in S$, and there does not exist $a \in S$, such that $\{a,b\} \in h'$ blocks (h,v). Towards a contradiction that for some $\{a,b\} \in h'$ with $a \in S$, we have $v'(b) > v(b)$. Then since $v'(a) > v(a)$, it turns out that $\{a,b\}$ blocks (h,v) contrary to hypothesis. This proves the lemma. Q.E.D.

Hence if in Lemma 2, (h,v) is a stable matching, then v'(a) > v(a) for all a∈S implies [v(b) ≥ v'(b) for all b∈μ'(S)], where μ' is the matching associated with h'.

4. Existence of Stable Outcomes for a two-sided contract choice problem:

Let F and W be two non-empty disjoint subsets of X such that F ∪ W = X. F is a set of firms and W is a set of workers.
A room-mates problem G is said to be a two-sided contract choice problem if for all f,f'∈F and w,w'∈W with f ≠ f' and w ≠ w': G({f,f'}) = G({w,w'}) = ϕ.
If (h,v) is an outcome for a two-sided contract choice problem, then S∈h implies #S ∈ {1,2}. Further, if #S = 2, then S∩F ≠ ϕ and S∩W ≠ ϕ.
In this section we shall denote an outcome (h,v) of a two-sided contract choice problem G by (μ,v), where μ is the matching associated with h. Clearly for all f∈F, μ(f) ∈ W∪{f} and for all w∈W, μ(w) ∈ F∪{w}.

Theorem 1: Every two-sided contract choice problem admits a stable outcome.

Proof: Let G be a given two-sided contract choice problem, and let f∈F and w∈W. Let $W^*(f) = \bigcup_{w' \in W}(\{w'\} \times G(\{f,w'\})) \cup \{(f, 0)\}$ and $F^*(w) = \bigcup_{m' \in M}(\{f'\} \times G(\{f',w\})) \cup \{(w, 0)\}$.
f has preferences defined by a binary relation $\geq_f$ over $W^*(f)$ satisfying the following property: for all (a,x), (b, x') ∈ $W^*(f)$: (a, x) $\geq_f$ (b, x') if and only if x(f) ≥ x'(f). Similarly, w has preferences defined by a binary relation $\geq_w$ over $F^*(w)$ satisfying the following property: for all (a, x), (b, x') ∈ $F^*(w)$ : (a, x) $\geq_w$ (b, x') if and only if x(w) ≥ x'(w).
Let $>_f$ denote the asymmetric part of $\geq_f$ and $>_w$ denote the asymmetric part of $\geq_w$
Let $W^{**}(f) = \{(w',x) \in W^*(f) / (w',x) >_f (f,0)\}$ and $F^{**}(w) = \{(f',x) \in F^*(w) / (f',x) >_w (w,0)\}$.
Given f ∈F and an element A of $W^{**}(f)$, let $A_{|W} = w'$, where (w', x) = A. Given w∈W and an element A of $F^{**}(w)$, let $A_{|F} = f'$, where (f', x) = A. Given a subset S of $\bigcup_{m \in M} W^*(m) \cup \bigcup_{w \in W} F^*(w)$, f∈F and w∈W, let U(f, S) = {(a, x')∈S ∩W*(f) / there does not exist (b,x")∈S: (b, x") $>_f$ (a, x')} and U(w,S) = {(a, x')∈S∩ F*(w)/ there does not exist ( b, x")∈S: (b,x") $>_w$ (a , x')}.
Let $F^1$ = {f∈F/ $W^{**}(f)$ ≠ϕ}. For f∈$F^1$, let $P^1(f)$ ∈ U(f, $W^{**}(f)$) where (w,x) = $P^1(f)$ implies f proposes to w the division in G({f,w}) where f gets x(f) and w gets x(w). Each f∈$F^1$ proposes to the worker $P^1(f)_{|W}$. For w∈ {$P^1(f)$/ f∈$F^1$}, let $R^1(w)$ = {(f, x) / (w, x)= $P^1(f)$}, $R^1_+(w) = R^1(w) \cap F^{**}(w)$ and $E^1(w)$ be any element of U(w, $R^1_+(w)$). Each w receiving a proposal, rejects all proposals in $R^1(w)$ \ {$E^1(w)$}. The proposal $E^1(w)$ is kept attached by w. Only those firms who are not kept attached at this step, are allowed to propose at the subsequent stage.

Suppose that the procedure continues to a stage 'k', $k \geq 1$, with $F^k$, $P^k(f)$ for $f \in F^k$, $R^k(w)$, $R^k_+(w)$ and $E^k(w)$ for $w \in \{P^k(f)/ f \in F^k\}$ having been defined. The procedure stops if $F^{k+1} = \{f \in F^1/$ all the proposals made by 'f' at the previous step were rejected and $W^{**}(f) \setminus \bigcup_{j=1}^{k}\{P^j(f)\} \neq \phi\} = \phi$. If $F^{k+1} \neq \phi$, then for $f \in F^{k+1}$, let

$P^{k+1}(f) \in U(f, W^*(f) \setminus \bigcup_{j=1}^{k} P^j(f))$. Each $f \in F^{k+1}$ proposes to the worker in $P^{k+1}(f)_{|W}$.
If $(w,x) = P^{k+1}(f)$, then $f$ proposes to $w$ the division where $f$ gets $x(f)$ and $w$ gets $x(w)$. For $w \in \bigcup_{m \in M^{k+1}}\{P^{k+1}(f)_{|W}\}$, let $R^{k+1}(w) = \{(f, x) / (w, x) = P^{k+1}(f)\}$ and
$R^{k+1}_+(w) = R^{k+1}(w) \cap F^{**}(w)$. Let $E^{k+1}(w)$ be any element of $U(w, E^k(w) \cup R^{k+1}_+(w))$. The proposal $E^{k+1}(w)$ is kept attached by $w$ at this step. The remaining proposals in $\{E^k(w)\} \cup R^{k+1}(w)$ are rejected.
Since $F \cup W$ is finite, there exists a stage $K$ when $F^K = \phi$. At this stage every $f \in F^1$ is either attached to some worker or has been rejected by every worker in $W^{**}(f)$. Further, every worker $w \in W$ for whom $F^{**}(w) \neq \phi$ has either not received any proposal or is attached to a firm.
Define an outcome $(\mu, v)$ as follows: for all $a \in \{f \in F/ W^{**}(f) = \phi\} \cup \{w \in W/ F^{**}(w) = \phi\}$, let $\mu(a) = a$ and $v(a) = 0$. For all $w \in W$, who never received a proposal or rejected each and every that she received, let $\mu(w) = w$ and $v(w) = 0$. For all $f \in F$, who have been rejected by every worker it has proposed to let $\mu(f) = f$ and $v(f) = 0$. The remaining workers are the ones who are attached at stage $K$. If $E^K(w) = (f, x)$, then let $(\mu(w), v(w)) = (f, x(w))$ and $(\mu(f), v(f)) = (w, x(f))$. Suppose there exists a pair $(f,w) \in F \times W$ such that $(f,w)$ blocks $(\mu, v)$. Thus, there exists $x \in G(\{f,w\})$ such that $x(f) > v(f)$ and $x(w) > v(w)$. Thus, $(w, x) >_f (\mu(f), (v(f), v(\mu(f))))$. Thus, $f$ must have proposed $(w, x)$ to $w$ and was rejected by $w$ in favor of some other proposal before it proposed $(\mu(f), (v(f), v(\mu(f))))$ to $\mu(f)$. Since $\geq_w$ is transitive, it must be the case that $(\mu(w), (v(w), v(\mu(w)))) \geq_w (f, x))$. This contradicts $x(w) > v(w)$ and proves the theorem. Q.E.D.

Let O be the set of outcomes of the procedure defined in the proof of Theorem 1. Clearly O though non-empty and finite can admit more than one element. An immediate consequence of the procedure, used in the proof of Theorem 1, is the following result.

Proposition 1: **_Weak Pareto Optimality for Firms: Let $(\mu^*, v^*) \in O$. Then, there does not exist any outcome which every firm prefers to $(\mu^*, v^*)$._**

Proof: If $\# F > \# W$, then there is no way in which the proposition can be falsified, since in every matching some firm must be without a worker. On the other hand, every worker who is single at $(\mu^*, v^*)$, continues to remain so at any other matching, where all firms are better off. This is because, according to the procedure defined in Theorem 1, a worker who is single, either rejected all the proposals she received preferring to remain single, or every firm considers its outcome at $(\mu^*, v^*)$ to be at least as good as any allocation that is feasible when it is paired with this worker. Hence, we can assume that $\mu^*$ maps F onto W, and in particular $\#F = \# W$.

Towards a contradiction suppose there is an outcome $(\mu, v)$ such that $v(f) > v^*(f)$ for all $f \in F$. This in particular implies that $\mu$ maps F onto W. Let $f^*$ be the firm whose proposal was accepted at the last stage of the procedure defined in Theorem 1. Let $w^*$ be the worker who accepted its offer. If $(w^*, v(f^*), v(w^*))$ was the only offer that $w^*$ had received, then $(w^*, v(f), v(w^*))$ could not have been preferred to $(\mu^*(f), v(f), v(\mu^*(f)))$ by any $f \neq f^*$. Thus, there could be no firm to whom $(w^*, v(f), v(w^*))$ could be assigned under $(\mu,v)$ leading to an improvement for it over $(\mu^*, v^*)$. Thus, there must have been some other proposal $(w^*, x)$ made by an $f \neq f^*$, which was rejected by $w^*$ in favor of $(f^*, v(w^*), v(f^*))$. Hence, f is assigned no worker under the $\mu^*$, contradicting that $\mu^*$ maps F onto W. This proves the proposition. Q.E.D.

The implication of Proposition 1 is that given any stable outcome in O, there is no outcome that all firms prefer to the former. However, there may exist outcomes (in fact stable outcomes!), where some firms are better off than at the earlier stable outcome, while others obtain the same pay-off as in the former. This possibility is observed in an illustration that we provide after an observation following Proposition 2.

A generalized two-sided matching problem G is said to be pair-wise efficient, if for all $(f,w) \in F \times W$ and $x, x' \in G(\{f,w\})$: $[x(f) > x'(f)]$ if and only if $[x(w) < x'(w)]$.

Note: If G is a two-sided contract choice problem satisfying pair-wise efficiency then given any two stable outcomes $(\mu, v)$ and $(\mu', v')$ for G: $[v'(f) \geq v(f)$ for all $f \in F]$ implies $[[v(w) \geq v'w)$ for all $w \in W]$. For if $v'(w) > v(w)$ for some $w \in W$, then $\mu'(w) \in F$ and by Lemma 2, $v(\mu'(w)) \geq v'(\mu'(w))$. Thus, $v(\mu'(w)) = v'(\mu'(w))$, contradicting pair-wise efficiency.

Another consequence of the procedure used to prove Theorem 1, is the following:

Proposition 2: ***Existence of Optimal Stable Outcome for Firms*** : Suppose *G is efficient and that for all $f \in F$ and $w,w' \in W$ with $w \neq w'$: $\{x(f)/ x \in G(\{f,w\})\} \cap \{x(f)/x \in G(\{f,w'\})\} = \phi$. Let $(\mu^*, v^*) \in O$. Then, O is singleton and given any stable outcome $(\mu, v)$: $v^*(f) \geq v(f)$ for all $f \in F$.*

Proof: Suppose that at the first step of the procedure, a firm f has a proposal $(w,v(f),v(w))$ rejected by a worker $w = \mu(f)$. Since, worker w, proceeds up her ranking during the procedure, w rejects $(w,v(w))$ of f in favor of some other proposal $(w,x)$ made by some other firm f'. Thus, $x(w) > v(w)$. Since, f' made the offer to w at the very first step, $x(f') \geq v(f')$. Now, $x \in G(\{f',w\})$ and $x(f') = v(f')$ implies by our assumption on G that $\mu(f') = w$. This combined with x, $(x(f'),v(w))$

∈G({f',w}) and x(w) > v(w), contradicts the pair-wise efficiency of G. Thus, x(f')
> v(f') and x(w) > v(w) contradicting the stability of (μ,v).
Suppose that up to a certain stage in the procedure, no firm f has a proposal
(w,v(f),v(w)) rejected by a worker w = μ(f). Suppose that at the next stage of the
procedure, a firm f has a proposal (w,v(f),v(w)) rejected by a worker w = μ(f) in
favor of another proposal (w,x) by another firm f'. Thus, x(w) > v(w). By the
induction hypothesis, f' has not had (μ(f'), v(f'), v(μ(f'))) by μ(f') up to the stage,
where it makes the offer to w in the procedure. Since, f' moves one rank down at a
time in the procedure, x ≥ v(f'). Now, x ∈G({f',w}) and x(f') = v(f') implies by our
assumption on G that μ(f') = w. This combined with x, (x(f'),v(w)) ∈G({f',w})
and x(w) > v(w), contradicts the pair-wise efficiency of G. Thus, x(f') > v(f') and
x(w) > v(w) contradicting the stability of (μ,v). Thus, even at this stage of the
procedure, no firm f has a proposal (w,v(f),v(w)) rejected by a worker w = μ(f).
Since the procedure terminates in a finite number of steps, it must be the case that
v*(f) ≥ v(f) for all f∈F.
Suppose (μ, v) and (μ', v') ∈ O. Thus, v(f) = v'(f) for all f∈F. If for some f∈F, μ(f)
≠ μ'(f), then {x(f)/ x∈ G({f,μ(f)})}∩{x(f)/x∈G({f,μ'(f)})} ≠ φ, contradicting our
requirement on G. Thus, μ(f) = μ'(f). By pair-wise efficiency of G, v(μ(f)) =
v'(μ'(f)). Thus, O is singleton. Q.E.D.

Note: G({f,w})∩G({f,w'}) = φ implies {x(f)/ x∈ G({f,w})}∩{x(f)/x∈G({f,w'})}
= φ. Thus, the consequences of Proposition 2 would continue to be valid if for all
f∈F and w,w'∈W with w≠ w': G({f,w})∩G({f,w'}) = φ.

The implication of Proposition 2 is the following: Suppose, a two-sided contract
choice problem is pair-wise efficient and is such that the set of yields that a firm
can achieve with any worker is disjoint from the set of yields it can achieve with
any other. Then there exists a stable outcome that is no worse for any firm
compared to the unique stable outcome in O.

Observation: If (μ,v) is any stable outcome for a two-sided contract choice
problem G satisfying the conditions of Proposition 2, then the set of firms that
employ workers at (μ,v) (: {f∈F/ v(f) > 0}) is a subset of the set of firms that
employ workers at (μ*,v*) (: {f∈F/ v*(f) > 0}). By the note preceding Lemma 2,
and the fact that v*(f) ≥ v(f) for all f∈F we have v(w) ≥ v*(w) for all w∈W. Thus,
the set of workers who are employed at (μ*,v*) (: {w∈F/ v*(w) > 0}) is a subset
of the set of workers who are employed at (μ,v) (: {w∈F/ v*(w) > 0}). The
number of workers who are employed at any stable outcome is equal to the
number of firms that employ them. Thus, the set of firms that employ workers is
the same for both stable outcomes and the set of workers who are employed are
also the same for both stable outcomes. Hence, given any two stable outcomes the
set of firms that employ workers and the set of workers who are employed are
both invariant.

As an illustration of the procedure used in the proof of Theorem 1, consider the situation where X = {1,2,3,4}, F = {1,2} and W = {3,4}. Let G be a two-sided contract choice problem defined thus:
G({1,3}) = {(x(1),x(3))/ (x(1),x(3)) = (3,1) or (x(1),x(3)) = (1,3)}
G({1,4}) = {(x(1),x(4))/ (x(1),x(4)) = (4,1) or (x(1),x(4)) = (1,4)}
G({2,3}) = {(x(2),x(3))/ (x(2),x(3)) = (3,2) or (x(2),x(3)) = (2,3)}
G({2,4}) = {(x(2),x(4))/ (x(2),x(4)) = (4,2) or (x(2),x(4)) = (2,4)}.

To begin with firm 1 proposes to worker 4, the division (x(1),x(4)) = (4,1) and firm 2 proposes to the worker 4, the division (x(2),x(4)) = (4,2). At this stage worker 3 has not received any offer. Worker 4, rejects the offer made by firm 1 since firm 2's offer is more attractive to worker 4. Worker 4, keeps firm 2's offer attached. At this stage firm 1 can either propose (x(1),x(3)) = (3,1) or (x(1),x(3)) = (1,3) to worker 3 or propose (x(1),x(3)) = (1,4) to worker 4. Of the three the most attractive offer to firm 1 is (x(1),x(3)) = (3,1), which firm 1 proposes to worker 3. In the absence of any competing offers, worker 3 keeps firm 1's offer attached. Since both firms and both workers are attached the matching $\mu$ where $\mu(1) = 3$, $\mu(2) = 4$ and the pay-off function v where $v(1) = 3$, $v(2) = 4$, $v(3) = 1$ and $v(4) = 2$ are defined. It is easily verified that $(\mu,v)$ is a stable outcome. Further, since firm 2, gets its highest possible pay-off, the outcome is Weakly Pareto Optimal for firms. It is worth noting that this is the only possible stable outcome for G. Since firm 2 gets its highest possible pay-off, it would never be a party to a blocking coalition. If {1,4} were to form blocking coalition then in order to give worker 4 more than $v(4) = 2$, firm 1 would have to settle for the pay-off of one unit which is less than $v(1) = 3$. Further, since every one gets a positive pay-off, none would prefer to remain single. Hence, the conclusions of Proposition 2 are (trivially!) verified.

However, if G({1,4}) = {(x(1),x(4))/ (x(1),x(4)) = (4,1) or (x(1),x(4)) = (3,3)}, while G satisfies pair-wise efficiency it fails to satisfy the condition that f∈F and w,w'∈W with w≠ w':{x(f)/ x∈ G({f,w})}∩{x(f)/x∈G({f,w'})} = φ, since 3∈{x(1)/ x∈ G({1,3})}∩{x(1)/x∈G({1,4})}. Now, after being rejected by worker 4, firm 1 could either proceed as before or propose (x(1),x(4)) = (3,3) to worker 4. In either case the pay-off to firm 1 is 3. In the latter situation worker 4, would reject the earlier offer and get attached to this new offer made by firm 1. The rejected firm 2, would now choose its best possible offer i.e. (x(2),x(3)) = (3,2) and propose it to worker 3. Worker 3 would get attached to this offer. Since both firms and both workers are attached the matching $\mu'$ where $\mu'(1) = 4$, $\mu'(2) = 3$ and the pay-off function v where $v(1) = 3$, $v(2) = 3$, $v(3) = 2$ and $v(4) = 3$ are defined. Since both the firms get 3 units each, and since it is impossible for both firms to get 4 units each, he outcome is Weakly Pareto Optimal for firms. Further neither {4,2} nor {1,3} is a blocking coalition for $(\mu',v')$. Hence $(\mu',v')$ is a stable outcome. Further, $(\mu,v)$ as defined earlier, continues to be a stable outcome:
(a) since firm 2 gets its highest possible pay-off, it would never be a party to a blocking coalition;

(b) if {1,4} were to form blocking coalition then in order to give worker 4 more than v(4) = 2, firm 1 would have to settle for the pay-off of three units which is no better than v(1) = 3.

It is easily observed that all the conclusions of Proposition 2 are violated.

5. Conclusion:

The main contribution of this paper is an extension of the classical two-sided matching market of Gale and Shapley (1962), now viewed as a two-sided contract choice problem. In this framework, we are able to prove that a stable outcome always exists. In the particular case where a two-sided contract choice problem reduces to a two-sided matching market, the procedure that we used to prove Theorem 1 reduces to the Deferred Acceptance Procedure with firms making offers (Gale and Shapley (1962)):

To start each firm makes an offer to its favorite worker, i.e. to its first worker on its list of acceptable workers. Each worker rejects the offer of any firm who is unacceptable to him, and each worker who receives one or more acceptable offers, rejects all but his most preferred of these. Any firm whose offer is not rejected at this point is kept "pending".

At any step any firm whose offer was rejected at the previous step, makes an offer to its next choice (i.e., to its most preferred acceptable worker, among those who have not rejected its offer), so long as there remains an acceptable worker to whom it has not yet made an offer. If at any step of the procedure, a firm has already made offers to, and been rejected by all workers it finds acceptable, then it makes no further offers. Each worker receiving offers rejects any from unacceptable firms, and also rejects all but his most preferred among the group consisting of the new offers together with any firm that he may have kept pending from the previous step.

The algorithm stops after any step in which no firm is rejected. At this point, every firm is either kept pending by some worker or has been rejected by every worker on its list of acceptable workers. The matching that is defined now, associates to each firm the worker who has kept him pending, if there be any. Further, workers who did not receive any offers at all and firms who have been rejected by all the workers on their list of acceptable workers, remain single.

In the above procedure, each firm, proceeds down its list of acceptable workers, and each worker proceeds up his list of acceptable firms.

Two major results that are available in the literature follow as a direct consequence of the deferred acceptance procedure. The first concerns the Weak Pareto Optimality of the matching defined by the procedure. Our Proposition 1 is an unconditional extension of that result to two-sided contract choice problems. The second result states that there is no other stable matching which a firm prefers to the one defined by the deferred acceptance procedure. Considering that a classical two-sided matching market satisfies pair-wise efficiency as well as the requirement that the set of yields that a firm can achieve with any worker is disjoint from the set of yields it can achieve with any other, Proposition 2 extends this result to the context of two-sided contract choice problems. A consequence of the two assumptions invoked for Proposition 2 is that the outcome of the procedure we define is unique. Hence, we arrive at the following stronger version of the existence result of a firm optimal stable matching for two-sided matching market (see Roth and Sotomayor (1990)):

Suppose a two-sided contract choice problem satisfies pair-wise efficiency and is such that the set of yields that a firm can achieve with any worker is disjoint from the set of yields it can achieve with any other. Then there exists a stable outcome that is no worse than any other stable outcome for any firm.

A major consequence of our analysis is the added flexibility that negotiation procedures in a labor market enjoy while conforming to the requirements for the existence of a stable outcome.

As markets and institutions evolve, newer opportunities become available on either side of a transaction thereby providing greater scope for designing more innovative contracts between firms and workers. Thus, with the passage of time the framework of two-sided contract choice problems is likely to prove more instrumental for the study and analysis of labor markets than its classical predecessor.

*Acknowledgment*: This paper is a revised version of an earlier working paper, Lahiri (2003). An earlier version of this paper was presented on January 9, 2004, at ICOR 2004, held at Indian Statistical Institute, Kolkata, India. I would like to thank Professor William Wallace for drawing my attention to a possible link between matching people in a market sense and the problem of contract choice.